\documentclass[12pt]{iopart}

\expandafter\let\csname equation*\endcsname\relax
\expandafter\let\csname endequation*\endcsname\relax
\usepackage{amsmath} 
\usepackage{graphicx}
\usepackage{subcaption}
\usepackage{float}
\usepackage{booktabs}

\usepackage[numbers,sort&compress]{natbib}

\begin{document}

\title{Quasi-single-stage optimization for advanced stellarators}

\author{Guodong Yu$^{1,\dagger}$, Yidong Xie$^{1,\dagger}$, Hengqian Liu$^1$, Caoxiang Zhu$^{1,*}$}
\address{$^1$ CAS Key Laboratory of Frontier Physics in Controlled Nuclear Fusion, School of Nuclear Science and Technology, University of Science and Technology of China, Hefei, Anhui 230026, China}

\footnotetext[1]{These authors contribute equally.}

\ead{caoxiangzhu@ustc.edu.cn}
\vspace{10pt}

\begin{abstract}
Advanced stellarator design requires a balance between plasma performance and the manufacturability of three-dimensional modular coils. In conventional two-stage optimization, the coils required to realize an optimized equilibrium can be limited by engineering feasibility.
Here, we develop a quasi-single-stage (QSS) framework that incorporates coil feasibility directly into plasma-boundary optimization. QSS uses the maximum normalized normal-field error, evaluated rapidly from surface currents on a uniformly offset winding surface, as a coil-feasibility surrogate. We apply this method to optimize configurations targeting quasi-axisymmetry, quasi-helical symmetry, quasi-isodynamicity, and a combination of omnigenity with piecewise omnigenity.
The QSS-optimized configurations exhibit smoother plasma boundaries and winding surfaces, lower normal-field reconstruction errors, and reduced coil complexity, while preserving favourable magnetic-symmetry and transport properties. 
QSS provides a practical proof-of-principle strategy for co-optimizing plasma physics and coil engineering in stellarator design.
\end{abstract}


\section{Introduction} 
\label{sec:introduction}

Advanced stellarators \cite{Wobig_advancedstellarator_1999} offer steady-state operation and reduced disruption risk, but their design must reconcile plasma confinement with the manufacturability of three-dimensional (3D) coils. A widely used strategy is nevertheless the conventional two-stage workflow, in which equilibrium optimization and coil design are largely separated. The first stage optimizes the target plasma. Dedicated stellarator optimization tools, including STELLOPT \cite{lazerson_stellopt_2020}, ROSE \cite{drevlak_rose_2019}, SIMSOPT \cite{Landreman_simsopt_2021}, and DESC \cite{dudt_desc_2020}, iterate the fixed-boundary equilibrium codes until desired properties are achieved. The second stage attempts to realize the target equilibrium with coils, using surface-current approaches like REGCOIL \cite{landreman_regcoil_2017} or three-dimensional filamentary approaches like FOCUS \cite{zhu_focus_2018} (SIMSOPT and DESC have FOCUS-like implementations). This separation keeps the problem modular and computationally efficient, and it has enabled successful optimized stellarator designs such as HSX~\cite{anderson_hsx_1995}, W7-X~\cite{klinger_w7x_2019}, and CFQS~\cite{liu_cfqs_2018}. However, it can also produce target equilibria that are difficult to realize accurately or economically. For NCSX \cite{zarnstorff_ncsx_2001}, the demanding modular-coil geometry was one of several engineering and programmatic factors that increased implementation difficulty and contributed to the termination of construction \cite{strykowsky2009Engineering}. MUSE, a recently constructed permanent-magnet stellarator, demonstrates a simpler hardware route, while also illustrating that discrepancies can remain between a fixed-boundary target and its free-boundary realization, including boundary error fields and edge magnetic islands \cite{qian_muse_2022}.

Single-stage optimization reduces this separation by combining equilibrium and coil-design objectives within one optimization problem. The optimized degrees of freedom can be coil variables, equilibrium variables, or both. Jorge \etal \cite{jorge_singlestage_2023} optimized plasma-boundary and coil-shape variables together, making the physics-engineering trade-off explicit in the objective function. Drevlak \etal \cite{drevlak_rose_2019} optimized only the Fourier coefficients of the equilibrium boundary, while using a surface-current method to obtain several coil-feasibility surrogate target quantities. Giuliani \etal \cite{giuliani__direct_2022} optimized only the Fourier coefficients of coil shapes and, in vacuum, computed flux surfaces in Boozer coordinates and the magnetic-field distribution on those surfaces directly from the coils; this coil-variable route was later extended in a global direct coil-design strategy \cite{giuliani_quasr_2024}. These approaches address the equilibrium-coil mismatch more directly than a purely two-stage workflow, but they also introduce two main practical difficulties. First, the number of optimization variables can grow rapidly, especially when both plasma-boundary and coil-shape variables are included, as in the formulation of Jorge \etal. Second, the objective function must balance many competing terms. These include symmetry, aspect ratio, rotational transform, field reconstruction, coil length, curvature, coil separation, and coil feasibility or complexity. Improving one term might therefore degrade another unless the weights and continuation strategy are carefully chosen. Together, the high-dimensional design space and the competition among objectives can make the optimization insufficiently robust and often require a warm start.

Quasi-single-stage (QSS) optimization follows this idea. It keeps the plasma boundary Fourier coefficients as the optimization variables, but adds coil-feasibility surrogates to the equilibrium objective. Thus, the optimizer receives engineering information without directly optimizing filamentary coils. A related QSS strategy was previously applied to permanent-magnet stellarators \cite{yu_qss_2024}, where REGCOIL-based surrogates helped reduce the number of permanent magnets while improving quasi-symmetry in MUSE++. Here we extend the idea to modular-coil stellarator design.

A useful QSS surrogate should be inexpensive to evaluate from the equilibrium, should not dominate the physics objectives, and should be relevant to the later coil-design problem. Existing surface-current quantities, such as \(K_{\mathrm{max}}\) and \(K\cdot\nabla K\), can represent coil spacing or curvature tendencies \cite{merkel_nescoil_1987}, while \(L_{\nabla B}\) estimates coil-plasma spacing directly from the equilibrium \cite{kappel_lgradb_2024}. There is also recent work that proposes several metrics for QSS optimization \cite{fu2026Flexible}. These quantities are valuable but target individual constraints. The present work instead uses the worst local field-reconstruction error as a compact surrogate for coil feasibility.

We use the normalized maximum normal field error from surface currents, \(b_n^{\mathrm{max}}\), as this surrogate. The winding surface is generated by a uniform outward-normal offset from the last closed flux surface (LCFS) and evolves with the plasma boundary during optimization. In this setting, \(b_n^{\mathrm{max}}\) measures the worst local field-reconstruction mismatch that surface currents or discrete coils must later correct. It is not a direct geometric measure of filamentary-coil complexity, but reducing it is expected to lower the burden on subsequent coil optimization under the controlled winding-surface construction.

The distinction from stand-alone post-processes with REGCOIL/NESCOIL is that \(b_n^{\mathrm{max}}\) and its gradient are evaluated during the boundary optimization and are used to update the plasma boundary, rather than only to assess coil feasibility after an equilibrium has been fixed. The implementation uses SIMSOPT with VMEC \cite{hirshman_vmec_1983} as the fixed-boundary equilibrium solver, OOPS \cite{liu_oops_2026} to quantify various symmetry targets, and REGCOIL \cite{landreman_regcoil_2017} to compute the surface-current surrogate. The aspect ratio and rotational transform constraints are included where needed to keep the optimized equilibria in the desired design class.
FOCUS \cite{zhu_focus_2018} is then used as a final check of the coil design. Within each symmetry class, coils for the reference and QSS-optimized equilibria are designed with equivalent procedures using FOCUS, so differences in field reconstruction and coil geometry can be interpreted as paired effects of the \(b_n^{\mathrm{max}}\) constraint. 

The paper is organized as follows. Section \ref{sec:method} introduces the essential parts of the optimization method, including OOPS, REGCOIL, and FOCUS. Section \ref{sec:QSS} describes the QSS optimization procedure for qa, qh, qi, and qi-pwO configurations. Section \ref{sec:results} compares equilibria and coils obtained with and without the \(b_n^{\mathrm{max}}\) surrogate. Section \ref{sec:conclusion} summarizes the main findings and limitations.

\section{Optimization Tools}
\label{sec:method}

Before introducing the actual QSS procedure, we briefly introduce the tools used. The main process is implemented in SIMSOPT, where OOPS evaluates symmetries of the magnetic configurations and REGCOIL supplies the coil-feasibility surrogate used during boundary optimization. After optimization, FOCUS designs modular coils for the final checks. This section gives the equations and implementation details of OOPS, REGCOIL, and FOCUS.

\subsection{OOPS}
\label{sec:oops}

OOPS \cite{liu_oops_2026} quantifies magnetic-field symmetry using the improved Cary--Shasharina (C--S) coordinate transformation \cite{cary_csmapping_1997}. It unifies quasi-axisymmetry (qa), quasi-helical symmetry (qh), quasi-isodynamicity (qi), and a combination of quasi-isodynamicity and piecewise-omnigenity (qi-pwO) \cite{liu2026Optimization, velasco2026Combination} into one framework by minimizing a symmetry-breaking metric in transformed coordinates. (Throughout this work, lowercase labels (``qa'', ``qh'', ``qi'', and ``qi-pwO'') denote classes of magnetic fields, whereas uppercase labels (``QA'', ``QH'', ``QI'', and ``QI-PWO'') refer to the corresponding specific configurations considered in this study.) Using OOPS avoids switching objective functions between symmetry classes and avoids direct evaluation of the adiabatic invariant \(\mathcal{J}\), which keeps the QSS optimization relatively inexpensive.
By constructing a mapping between the orthogonal coordinate system \((\alpha, \eta)\) and the Boozer coordinate system \cite{boozer_coordinate_1981} \((\theta_B, \zeta_B)\), abstract magnetic field symmetry breaking is converted into mathematically well-defined metrics suitable for direct numerical optimization. The governing expressions are summarized below.

First, the improved C–S transformation relates the field-line label to the magnetic-field contour label:
\begin{equation}
\zeta(\alpha, \eta) = \eta - D(\eta) \, S(\alpha, \eta)
\label{eq:oops_cs}
\end{equation}
where \(\alpha \in [0, 2\pi)\) denotes the field-line label, \(\eta \in [-\pi, \pi)\) the magnetic-field contour label (\(\eta = 0\) corresponds to \(B_{\rm min}\), \(\eta = \pm\pi\) to \(B_{\rm max}\)); \(S(\alpha, \eta)\) controls the geometric shape of the \(B\) contour, and \(D(\eta)\) determines the bounce distance between pairs of equal-field contours along a field line.

To preserve inherent stellarator symmetry, \(S(\alpha, \eta)\) and \(D(\eta)\) are expanded as Fourier series in the form of odd and even functions, respectively:
\begin{equation}
\begin{cases}
\displaystyle S(\alpha, \eta) = \sum_{m} s_{m} \sin\!\left[m\,y(\alpha, \eta)\right], \\[4pt]
\displaystyle D(\eta) = \pi - |\eta| + \sum_{n} d_{n} \cos\!\left[\left(n+\frac{1}{2}\right)\eta\right],
\end{cases}
\label{eq:oops_fourier}
\end{equation}
where \(s_{m}\) and \(d_{n}\) are Fourier coefficients, \(y(\alpha, \eta) = Y [\alpha\ \ \eta]^\mathrm{T}\) describes the helical character of the magnetic field, and \(Y\) is a row vector whose form is determined by the target symmetry type.


Following the paradigm of quasi-symmetry optimization, the objective function in OOPS corresponds to the minimization of the symmetry-breaking metric in the transformed coordinates:
\begin{equation}
f_{\rm omni} = \left[\sum_{m \neq 0,n} \left( \frac{B_{m,n}}{B_{0,0}} \right)^2\right]^{1/2},
\label{eq:oops_obj}
\end{equation}
where \(B(\alpha, \eta) = \sum_{m,n} B_{m,n} \cos(m\alpha - n\eta)\) is the Fourier representation of the magnetic field in \((\alpha, \eta)\) coordinates, \(B_{m,n}\) are the corresponding Fourier coefficients, and \(B_{0,0}\) is the leading-order amplitude. For \(S = 0\), the objective reduces to the classical quasi-symmetry form and can be directly applied to \textit{qa} and \textit{qh} optimization.

Within the present work, OOPS is tightly coupled with the SIMSOPT framework and the VMEC fixed-boundary equilibrium solver. The optimization procedure is adopted for 2-period \textit{qa}, 3-period \textit{qi} and \textit{qi-pwO}, and 4-period \textit{qh}. First, the row vector \(Y\)  is assigned according to the helical characteristics of each configuration. Second, the Fourier coefficients \(s_m\) and \(d_n\) are fixed to avoid introducing unnecessary optimization variables: \(S = 0\) for \textit{qa} and \textit{qh}, \(S = 0.3\sin y\) for \textit{qi} and \textit{qi-pwO}, and \(D = \pi - |\eta|\) for all configurations. Finally, the total optimization objective is constructed using \(f_{\rm omni}\) from Eq.~\eqref{eq:oops_obj} as the central physical term, combined with aspect ratio and rotational transform constraints. 

\subsection{REGCOIL}
\label{sec:regcoil}

REGCOIL \cite{landreman_regcoil_2017} extends NESCOIL by adding Tikhonov regularization \cite{tikhonov_regularization_1963} to the surface-current inverse problem. Whereas NESCOIL minimizes the normal field error without penalizing the current distribution, REGCOIL balances field accuracy against current smoothness, making it useful for constructing coil-feasibility diagnostics from a target equilibrium.

The theoretical framework of REGCOIL is grounded in the surface current potential model. The surface current density \(\boldsymbol{K}\) on the winding surface is related to the current potential \(\Phi\) via
\begin{equation}
\boldsymbol{K} = \boldsymbol{n}' \times \nabla' \Phi,
\label{eq:regcoil_K}
\end{equation}
where \(\boldsymbol{n}'\) denotes the unit normal vector of the predefined winding surface, and \(\nabla'\) represents the surface gradient operator on the winding surface. The regularized objective function takes the form
\begin{equation}
\chi^2 = \chi_B^2 + \lambda \chi_K^2,
\label{eq:regcoil_total}
\end{equation}
with \(\lambda\) serving as the regularization parameter that balances magnetic-field-fitting accuracy and current-distribution smoothness. The normal field error integral \(\chi_B^2\) is defined as
\begin{equation}
\chi_B^2 = \iint_S (\boldsymbol{B}\cdot\boldsymbol{n})^2 dS,
\label{eq:regcoil_chiB}
\end{equation}
where \(S\) is the plasma surface, \(\boldsymbol{B}\) is the total magnetic field on \(S\), and \(\boldsymbol{n}\) is the unit normal vector of \(S\). For the vacuum configurations considered here, this field is generated by the optimized surface current on the winding surface. The current density regularization term is given by
\begin{equation}
\chi_K^2 = \iint_{S'} |\boldsymbol{K}|^2 dS',
\label{eq:regcoil_chiK}
\end{equation}
where \(S'\) denotes the winding surface. 

In the present work, the dimensionless maximum normal field error, normalized by the local magnetic field magnitude, is adopted as the primary coil-feasibility surrogate objective:
\begin{equation}
b_n^{\mathrm{max}} = \max_{\boldsymbol{r}\in S} \frac{|\boldsymbol{B}(\boldsymbol{r})\cdot\boldsymbol{n}(\boldsymbol{r})|}{B(\boldsymbol{r})},
\label{eq:regcoil_Bnmax}
\end{equation}
where \(B(\boldsymbol{r}) = |\boldsymbol{B}(\boldsymbol{r})|\) represents the local magnetic field strength at position \(\boldsymbol{r}\).
$b_n^{\mathrm{max}}$ penalizes the discrepancy between the target field and the achieved field produced by surface currents.
Surface currents generally have lower normal field errors than filamentary coils.
So if $b_n^{\mathrm{max}}$ is large, it usually means that the equilibrium cannot be easily achieved by coils.
In this work, the winding surface is uniformly expanded from the plasma boundary, which is a generally good choice in surface-current approaches.
The synchronization between the plasma boundary and winding surface will also help avoid extreme solutions like self-intersecting equilibria.
Therefore, $b_n^{\mathrm{max}}$ is a good surrogate for coil feasibility, which we will demonstrate later. 


For the four configurations investigated--2-period \textit{qa}, 3-period \textit{qi} and \textit{qi-pwO}, and 4-period \textit{qh}--the implementation of REGCOIL follows a consistent procedure. All configurations are scaled to a major radius \(R_0 = 1.0\,\mathrm{m}\). The winding surface is constructed by offsetting the last closed flux surface outward along the normal direction: an offset distance of \(0.18\,\mathrm{m}\) is applied for \textit{qa}, \textit{qi} and \textit{qi-pwO} with an aspect ratio of \(6.0\), while \(0.15\,\mathrm{m}\) is used for \textit{qh} with an aspect ratio of \(8.0\). During the optimization, \(\lambda\) is fixed at \(0\), and the surface current distribution is solved to compute \(b_n^{\mathrm{max}}\). 

\subsection{FOCUS}
\label{sec:focus}

FOCUS \cite{zhu_focus_2018} represents each modular coil as a closed smooth space curve and optimizes the coil geometry without requiring a preset winding surface. It opens a new avenue for stellarator coil design, and related capabilities have been implemented in other codes such as SIMSOPT and DESC. Here, FOCUS is used only after equilibrium optimization to verify whether QSS can simplify the resulting coils. The coil curves are parametrized by Fourier series in Cartesian coordinates:
\begin{equation}
\begin{cases}
\displaystyle x(t) = x_{c,0} + \sum_{n=1}^{N_\mathrm{F}} \left[ x_{c,n}\cos(nt) + x_{s,n}\sin(nt) \right], \\[4pt]
\displaystyle y(t) = y_{c,0} + \sum_{n=1}^{N_\mathrm{F}} \left[ y_{c,n}\cos(nt) + y_{s,n}\sin(nt) \right], \\[4pt]
\displaystyle z(t) = z_{c,0} + \sum_{n=1}^{N_\mathrm{F}} \left[ z_{c,n}\cos(nt) + z_{s,n}\sin(nt) \right],
\end{cases}
\label{eq:focus_geometry}
\end{equation}
where \(t\in[0,2\pi]\) is the curve parameter, \(N_\mathrm{F}\) is the Fourier truncation mode, and \(\{x_{c,0},y_{c,0},z_{c,0}\}\) and \(\{x_{c,n},x_{s,n},y_{c,n},y_{s,n},z_{c,n},z_{s,n}\}\) serve as the geometric optimization variables.
The local curvature derived from this parametrization is
\begin{equation}
\kappa(t) = \frac{\| \boldsymbol{x}'(t) \times \boldsymbol{x}''(t) \|}{\| \boldsymbol{x}'(t) \|^3}.
\label{eq:focus_curve}
\end{equation}

Following the original FOCUS formulation, magnetic-field accuracy is quantified by the squared relative normal-field objective on the LCFS,
\begin{equation}
f_B(\boldsymbol{x}) = \int_S \frac{1}{2}
\left[
\frac{\boldsymbol{B}(\boldsymbol{x})\cdot\boldsymbol{n}}{|\boldsymbol{B}(\boldsymbol{x})|}
\right]^2 \mathrm{d}S,
\label{eq:focus_normal_field}
\end{equation}
where \(S\) is the LCFS and \(\boldsymbol{B}\) is the magnetic field produced by the coils for the vacuum configurations considered here. 
The FOCUS results reported below use the dimensionless area-averaged absolute relative normal-field error \(\left\langle |b_n| \right\rangle_S = \bigl(\int_S |\boldsymbol{B}\cdot\boldsymbol{n}|/|\boldsymbol{B}|\,\mathrm{d}S\bigr) / \bigl(\int_S \mathrm{d}S\bigr)\).
This diagnostic is distinct from the squared objective \(f_B\) used in the FOCUS optimization in Eq.~\eqref{eq:focus_normal_field}.
The physical and engineering objectives are combined in a weighted target function,
\begin{equation}
\chi^2(\boldsymbol{x}) = w_B f_B(\boldsymbol{x}) + \sum_{j\in\mathcal{E}} w_j p_j(\boldsymbol{x}),
\label{eq:focus_obj}
\end{equation}
with the dimensionless soft-target penalties
\begin{equation}
p_j(\boldsymbol{x}) = \left[
\frac{f_j(\boldsymbol{x})-f_{j,o}}{f_{j,o}}
\right]^2.
\label{eq:focus_soft_target}
\end{equation}
Here \(\boldsymbol{x}\) contains the coil parameters, \(\mathcal{E}\) denotes the set of engineering objectives, \(f_j(\boldsymbol{x})\) is the corresponding geometric quantity, \(f_{j,o}\) is its soft target, and \(w_B\) and \(w_j\) are weights. In this study, \(\mathcal{E}\) includes average coil length, maximum curvature, minimum coil--coil distance, and minimum coil--plasma distance. These target values guide the optimization but are not enforced as hard bounds.

Three-dimensional modular coils are designed separately for each symmetry class. Within a given class, the reference and QSS-optimized equilibria use identical FOCUS procedures and final soft-target settings. Across classes, the coil number and soft-target settings differ because the configurations have different topology, geometry, and convergence behavior.
All runs start from toroidally distributed circular coils. 
The \textit{qa}, \textit{qi} and \textit{qi-pwO} cases have four coils per half-period, while the \textit{qh} case has six coils per half-period.
Coil currents remain identical.

The coil optimization minimizes the normal-field objective on the LCFS while using soft targets for average coil length, maximum curvature, coil--coil distance, and coil--plasma distance. The numerical values reported here are the final soft targets used in the converged FOCUS runs, not values prescribed \textit{a priori}. For this non-convex engineering optimization, they were adjusted manually. Once a converged setting was identified for one case, the same coil initialization, optimization procedure, final soft targets, and Fourier order were applied to the paired reference and QSS-optimized equilibria. The FOCUS solutions are therefore used as matched, case-wise validation results rather than unique global optima or absolute manufacturability certifications. The final average-length targets for \textit{qa}, \textit{qh}, \textit{qi}, and \textit{qi-pwO} are \(5.00\,\mathrm{m}\), \(2.10\,\mathrm{m}\), \(4.0\,\mathrm{m}\), and \(3.50\,\mathrm{m}\); the maximum-curvature targets are \(9.74\,\mathrm{m}^{-1}\), \(9.6\,\mathrm{m}^{-1}\), \(9.2\,\mathrm{m}^{-1}\), and \(9.0\,\mathrm{m}^{-1}\); the coil--coil-distance targets are \(0.116\,\mathrm{m}\), \(0.06\,\mathrm{m}\), \(0.034\,\mathrm{m}\), and \(0.05\,\mathrm{m}\); and the coil--plasma-distance targets are \(0.31\,\mathrm{m}\), \(0.12\,\mathrm{m}\), \(0.13\,\mathrm{m}\), and \(0.158\,\mathrm{m}\). These class-specific soft targets define paired comparisons within each class rather than rankings across classes. The coil Fourier order is \(N_\mathrm{F}=4\) for all cases.

\section{Quasi-single-stage optimization}
\label{sec:QSS}

\subsection{Optimization framework}
\label{sec:optimization_framework}
 
The QSS workflow is implemented in SIMSOPT using VMEC, OOPS, and REGCOIL. Only the plasma boundary is optimized; coil geometry is not included as an optimization variable.

The sole optimization degrees of freedom are the Fourier coefficients \(R_{mnc}\) and \(Z_{mns}\) that describe the plasma boundary shape in VMEC, adhering to standard stellarator symmetry:
\begin{equation}
\begin{aligned}
R(\theta,\phi) &= \sum_{m,n} R_{mnc} \cos(m\theta - n_{\text{fp}} n \phi), \\
Z(\theta,\phi) &= \sum_{m,n} Z_{mns} \sin(m\theta - n_{\text{fp}} n \phi),
\end{aligned}
\end{equation}
where \(\theta\) denotes the poloidal angle, \(\phi\) the toroidal angle, and \(n_{\text{fp}}\) the number of field periods.
At each iteration, VMEC solves the equilibrium, OOPS evaluates the symmetry metric, and REGCOIL computes \(b_n^{\mathrm{max}}\). A trust-region reflective algorithm then updates the boundary Fourier coefficients.

The total optimization objective is a weighted sum of magnetic equilibrium objectives and the coil-feasibility surrogate objective:
\begin{equation}
f_{\text{total}} = \omega_{\text{omni}} f_{\text{omni}}^2 + \omega_{A_p} f_{A_p}^2 + \omega_{\iota} f_{\iota}^2 + \omega_{b_n^{\mathrm{max}}} f_{b_n^{\mathrm{max}}}^2,
\end{equation}
where \(\omega_{\text{omni}}\), \(\omega_{A_p}\), \(\omega_{\iota}\), and \(\omega_{b_n^{\mathrm{max}}}\) are weighting coefficients designed to prevent dominance by any single objective.
The magnetic symmetry objective $f_{\text{omni}}$, computed by OOPS, is reduced to be \( < 10^{-4}\) to meet the adopted symmetry stopping criterion.
The aspect ratio \(A_p = R_0/a\) (where \(R_0\) is the major radius and \(a\) the minor radius) governs device compactness and engineering cost. It is constrained to a target value to avoid excessive plasma boundary expansion or contraction:
\begin{equation}
f_{A_p} = A_p - A_p^*,
\end{equation}
where \(A_p^* = 6.0\) for \textit{qa}, \textit{qi} and \textit{qi-pwO} configurations, and \(A_p^* = 8.0\) for \textit{qh}.
The rotational transform \(\iota\) is a critical parameter for MHD stability, constrained to avoid low-order rational surfaces and associated magnetic islands:
\begin{equation}
f_{\iota} = \iota - \iota^*,
\end{equation}
where \(\iota\) is the rotational transform at the target flux surface, and \(\iota^*\) is the target value. Optimization focuses exclusively on the LCFS, with a strict target \(\iota^* = 0.4\) applied only for qa.
The coil-related objective is the dimensionless maximum normal field error \(b_n^{\mathrm{max}}\) from REGCOIL:
\begin{equation}
f_{b_n^{\mathrm{max}}} = b_n^{\mathrm{max}}.
\end{equation}
The condition \(f_{b_n^{\mathrm{max}}} < 10^{-4}\) is used as the coil-feasibility stopping threshold.

Throughout optimization, the weights for the normalized physical objectives are fixed at \(\omega_{\text{omni}} = 1\), \(\omega_{A_p} = 1\), and \(\omega_{\iota} = 1\). The coil-objective weight \(\omega_{b_n^{\mathrm{max}}}\) is treated as a tunable scaling coefficient and is selected according to the magnitude of \(b_n^{\mathrm{max}}\), so that the coil-feasibility term remains numerically active without overwhelming the physical objectives. Order-of-magnitude variations of this scaling produced the same qualitative optimization trends in trial calculations; consequently, no unique universal weight is implied.

\subsection{Optimization procedure}
\label{sec:optimization_procedure}
The four symmetry classes are optimized with the same staged procedure.
All configurations are scaled to \(R_0=1\,\text{m}\) and \(B_0=1.0\,\text{T}\). 
The magnetic symmetry optimization is only enforced on the LCFS.

\textbf{Phase I: Seed equilibrium generation}
The first phase generates a stable seed equilibrium using only magnetic symmetry, aspect ratio, and rotational transform objectives. The coil-feasibility surrogate is not included. Boundary modes are limited to \(m=1\) and \(n=1\), producing a low-dimensional rotating elliptical equilibrium.
This seed equilibrium, while not accounting for coil engineering feasibility, satisfies key physical parameter constraints and provides a stable initial configuration for subsequent refined optimization. Target rotational transforms are set to \(\iota^*=0.4\) for \textit{qa}, \(\iota^*=0.6\) for \textit{qi} and \textit{qi-pwO}, and \(\iota^*=-1.0\) for \textit{qh}. In subsequent high-resolution optimization, only \textit{qa} retains the rotational transform target.
The other configurations are released from this constraint because of their inherent symmetry topology.

\textbf{Phase II: Hierarchical iterative optimization}
Starting from the same seed, each symmetry class is split into two paired cases:
\begin{itemize}
    \item Group 1: Optimizes only physical objectives (magnetic symmetry, aspect ratio, and rotational transform) to further reduce the magnetic symmetry error.
    \item Group 2: Incorporates the coil-feasibility surrogate, \(b_n^{\mathrm{max}}\), into Group 1's objectives, enabling joint optimization of physical performance and engineering feasibility.
\end{itemize}
For a fair comparison, the two groups within the same symmetry class are initialized from the same seed equilibrium and use the same optimization algorithm, Fourier-mode continuation strategy, and numerical convergence tolerances. The physics-based stopping targets differ as described below. Their convergence trajectories and computational costs naturally differ because their objective functions differ. 
These quantities are not used as performance measures. The paired equilibria are compared at the same final plasma-boundary Fourier truncation, so they have the same boundary representational capacity, and the only difference in the optimized objective function is whether the \(b_n^{\mathrm{max}}\) surrogate is included.
For \textit{qi} and \textit{qi-pwO}, the magnetic field mapping on the weak-field side is identical. \textit{qi-pwO} does not optimize the strong-field side, while all other optimization procedures are the same as for \textit{qi}.

VMEC boundary Fourier modes \((m,n)\) are gradually increased starting from \((2,2)\). Group 1 is stopped when \(f_{\rm omni}<10^{-4}\). For Group 2, \(b_n^{\mathrm{max}}<10^{-4}\) is the primary stopping target, while \(f_{\rm omni}\) is monitored simultaneously and is also sought below \(10^{-4}\) when compatible with the coil-feasibility objective. The symmetry threshold is not imposed as a hard requirement for Group 2. In particular, the new configuration, QI+, does not reach it and is retained as a trade-off case. For inter-group comparability, Group 1's mode numbers are aligned with Group 2 using a maximum-mode strategy. Hence, both configurations of each pair are evaluated with the same final boundary truncation: \((4,4)\) for \textit{qa}, \((5,5)\) for \textit{qh}, and \((6,6)\) for \textit{qi} and \textit{qi-pwO}.

\textbf{Phase III: Coil design and validation}
FOCUS is applied to design coils for the equilibria in Phase-II. For each symmetry class, the reference and QSS-optimized configurations use identical coil initialization, final soft-target settings, Fourier truncation, and convergence settings. The objective minimizes the normal-field objective on the LCFS while applying soft targets to average length, maximum curvature, coil--coil distance, and coil--plasma distance.
Coil curvature, coil--coil clearance, and coil--plasma clearance are therefore included as soft engineering objectives during optimization. The truncated Fourier order for coil-shape parametrization is uniformly set to \(N_\mathrm{F}=4\) across all symmetry classes.




\section{Numerical results}
\label{sec:results}

\subsection{Overview of optimization results}
\label{sec:results_overview}
Eight configurations are obtained: reference and QSS-optimized cases for \textit{qa}, \textit{qh}, \textit{qi}, and \textit{qi-pwO}. The specific reference equilibria will be called QA, QH, QI, and QI-PWO, while the QSS-optimized equilibria will be QA+, QH+, QI+, and QI-PWO+. Table \ref{tab:equilibrium_params} summarizes the main equilibrium and REGCOIL quantities. This proof-of-principle study is restricted to fixed-boundary, zero-beta vacuum equilibria.
Finite-beta effects, bootstrap current, and MHD stability constraints are left for future work. The following subsections compare rotational transform, magnetic symmetry, neoclassical transport, boundary geometry, surface-current diagnostics, and FOCUS coil results.

\begin{table}[htbp]
  \centering
  \caption{Basic equilibrium parameters for all configurations. The values of \(b_n^{\mathrm{max}}\) are evaluated at \(\lambda=0\), whereas \(\chi_B^2\) and \(\chi_K^2\) are evaluated at the visually selected diagnostic points in the turning regions of the corresponding REGCOIL L-curves.}
  \label{tab:equilibrium_params}
  \resizebox{\linewidth}{!}{
  \begin{tabular}{lcccccccc}
    \toprule
    \textbf{Parameter} & \textbf{QA} & \textbf{QA+} & \textbf{QH} & \textbf{QH+} & \textbf{QI} & \textbf{QI+} & \textbf{QI-PWO} & \textbf{QI-PWO+} \\
    \midrule
    $N_{\rm fp}$ & 2 & 2 & 4 & 4 & 3 & 3 & 3 & 3 \\
    Aspect ratio & 6.0 & 6.0 & 8.0 & 8.0 & 6.0 & 6.0 & 6.0 & 6.0 \\
    Symmetry error & $8.5\times10^{-6}$ & $3.8\times10^{-5}$ & $8.7\times10^{-5}$ & $3.2\times10^{-4}$ & $1.0\times10^{-4}$ & $3.4\times10^{-2}$ & $3.1\times10^{-5}$ & $4.0\times10^{-5}$ \\
    Max-elongation & 6.9 & 3.5 & 4.5 & 2.5 & 6.6 & 6.1 & 5.4 & 5.1 \\
    Mean-elongation & 4.1 & 2.9 & 2.9 & 2.4 & 4.8 & 3.4 & 3.2 & 3.2 \\
    Minimum $L_{\nabla B}$ (m) & 0.57 & 0.54 & 0.23 & 0.27 & 0.23 & 0.35 & 0.36 & 0.40 \\
    $b_n^{\mathrm{max}}$ & $2.9\times10^{-4}$ & $7.3\times10^{-6}$ & $1.6\times10^{-4}$ & $4.6\times10^{-5}$ & $4.7\times10^{-3}$ & $8.8\times10^{-5}$ & $3.0\times10^{-4}$ & $4.6\times10^{-5}$ \\
    $\chi_B^2$ (T$^2 m^2$) & $6.5\times10^{-9}$ & $9.8\times10^{-12}$ & $5.6\times10^{-9}$ & $4.8\times10^{-10}$ & $2.0\times10^{-5}$ & $7.9\times10^{-9}$ & $5.9\times10^{-8}$ & $1.5\times10^{-9}$ \\
    $\chi_K^2$ (A$^2$) & $1.9\times10^{13}$ & $1.7\times10^{13}$ & $2.2\times10^{13}$ & $1.1\times10^{13}$ & $3.0\times10^{13}$ & $1.2\times10^{13}$ & $1.3\times10^{13}$ & $9.4\times10^{12}$ \\
    \bottomrule
  \end{tabular}
  }
\end{table}

\subsection{Rotational transform profile analysis}
\label{sec:rot_transform} 
\begin{figure*}[htbp]
  \centering
  \includegraphics[width=0.9\textwidth]{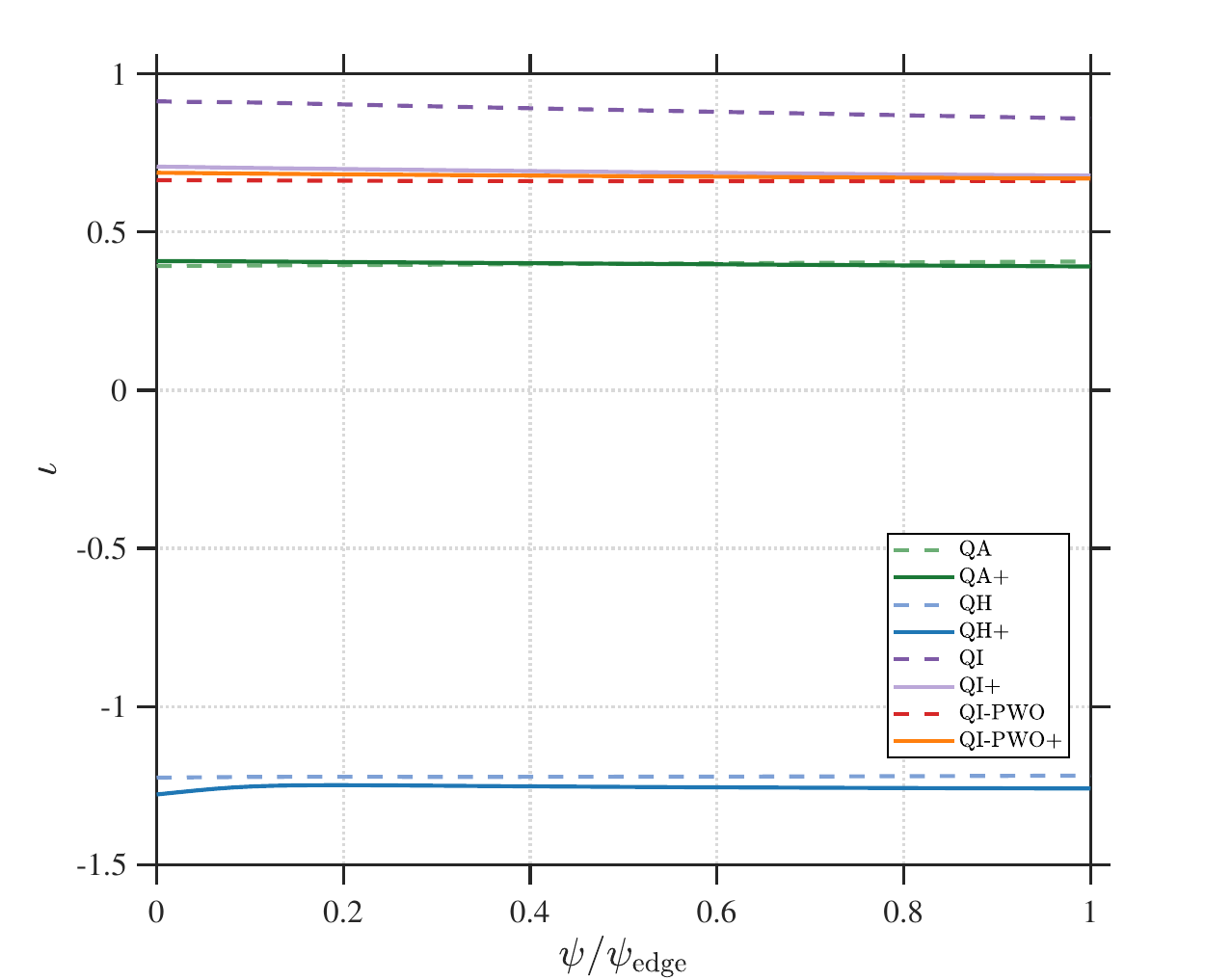} 
  \caption{Radial profiles of the rotational transform $\iota$ for all $8$ optimized configurations, plotted against the normalized toroidal magnetic flux $\psi/\psi_{\rm edge}$.}
  \label{fig:iota_profile}
\end{figure*}

Figure \ref{fig:iota_profile} shows the rotational-transform profiles for all eight configurations.
The QA and QA+ profiles are nearly identical, with \(\iota\) at the LCFS close to 0.4. The QH and QH+ profiles are also similar and retain the negative \(\iota\) expected for the \textit{qh} target. Thus, the \(b_n^{\mathrm{max}}\) constraint has little effect on the rotational-transform profiles of the \textit{qa} and \textit{qh} cases.
The profiles of QI and QI+ remain positive and relatively flat but differ in magnitude. QI reaches \(\iota\approx0.8\), whereas QI+ converges near the QI-PWO level, around 0.67. QI-PWO and QI-PWO+ are nearly identical, consistent with their shared flux-surface mapping and the partially-toroidal nature of the \textit{qi-pwO} target.

All eight configurations retain relatively flat rotational-transform profiles. The \(b_n^{\mathrm{max}}\) constraint mainly changes the accessible \(\iota\) level in the \textit{qi} class, rather than the intended field-period topology. The QI+ case also shows that a flat \(\iota\) profile alone does not guarantee high-quality omnigenity when a strong coil-feasibility surrogate is imposed.

\subsection{Magnetic field symmetry optimization and neoclassical transport analysis}
\label{sec:symmetry_neoclassical}

\begin{figure*}[htbp]
  \centering
  \includegraphics[width=\textwidth]{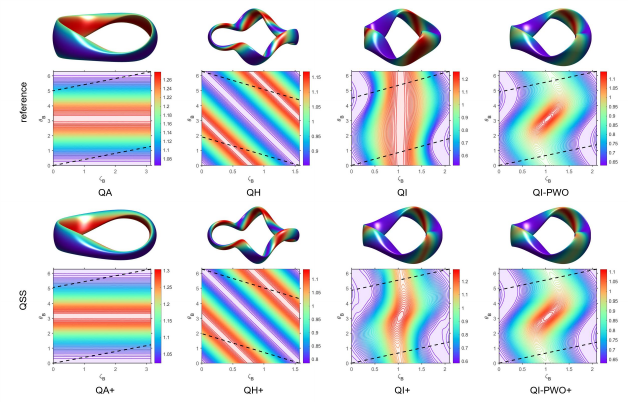}
  \caption{Three-dimensional plasma boundary shapes and magnetic-field strength in Boozer coordinates for all 8 optimized configurations. Columns correspond to the \textit{qa}, \textit{qh}, \textit{qi}, and \textit{qi-pwO} classes, and rows compare the reference and QSS-optimized configurations. Each entry stacks the boundary shape above the corresponding field-strength map in Boozer coordinates.}
  \label{fig:boozer_maps}
\end{figure*}

\begin{figure*}[htbp]
  \centering
  \includegraphics[width=0.9\textwidth]{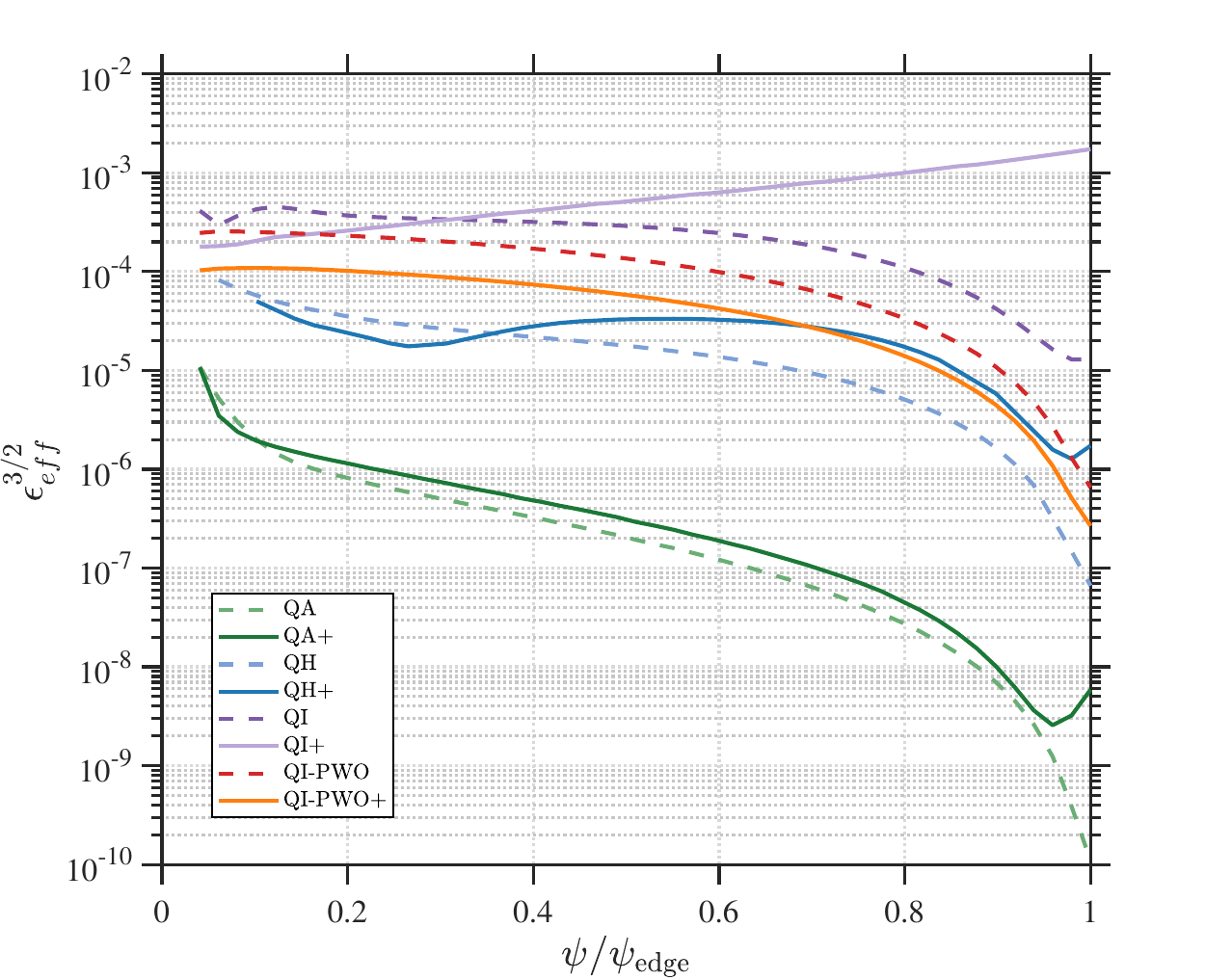} 
  \caption{Radial profiles of the effective neoclassical transport coefficient $\epsilon_{\rm eff}^{3/2}$ for all $8$ optimized configurations, plotted against the normalized toroidal magnetic flux $\psi/\psi_{\rm edge}$.}
  \label{fig:neoclassical_transport}
\end{figure*}

This section compares magnetic symmetry and neoclassical transport using the Boozer maps in Figure \ref{fig:boozer_maps}, the symmetry errors in Table \ref{tab:equilibrium_params}, and the \(\epsilon_{\rm eff}^{3/2}\) profiles in Figure \ref{fig:neoclassical_transport}. Here \(\epsilon_{\rm eff}^{3/2}\) is determined by the magnetic-field configuration and is proportional to the \(1/\nu\) neoclassical transport coefficient.

The reference cases all achieve low symmetry error: \(8.5\times10^{-6}\) for QA, \(8.7\times10^{-5}\) for QH, \(1.0\times10^{-4}\) for QI, and \(3.1\times10^{-5}\) for QI-PWO. The corresponding Boozer maps show the expected structures for magnetic symmetries.
Adding \(b_n^{\mathrm{max}}\) affects the four symmetry classes differently. QA+ and QH+ show only moderate symmetry degradation, with errors of \(3.8\times10^{-5}\) and \(3.2\times10^{-4}\). QI+ is the outlier: its symmetry error increases to \(3.4\times10^{-2}\), and the Boozer map is visibly distorted. By contrast, QI-PWO+ retains a low error of \(4.0\times10^{-5}\). This contrast indicates that the \textit{qi-pwO} target is more compatible with the present coil-feasibility surrogate than the standard \textit{qi} target in this test set.

The reference configurations achieve higher symmetry precision because their objective functions do not include the coil-feasibility surrogate, which competes with the other optimization objectives in the constrained cases. The constrained configurations expose the equilibrium optimizer to this coil-feasibility cost earlier. Except for QI+, the symmetry errors of QA+, QH+, and QI-PWO+ remain close to or below \(10^{-4}\). QI+ is therefore a trade-off case rather than a fully successful constrained \textit{qi} solution: the standard \textit{qi} target is difficult to combine with the present \(b_n^{\mathrm{max}}\) constraint without sacrificing symmetry quality. This negative example motivates the \textit{qi-pwO} route, for which QI-PWO+ preserves low symmetry error while satisfying the same coil-feasibility threshold.

The neoclassical transport trends follow the symmetry results. QA and QA+ have the lowest \(\epsilon_{\rm eff}^{3/2}\), approximately \(10^{-10}\) to \(10^{-5}\), with only a small edge increase in QA+. 
QH and QH+ are higher, approximately \(10^{-7}\) to \(10^{-4}\).
For the \textit{qi} class, QI+ has more than one order of magnitude larger transport coefficient than QI, consistent with its degraded symmetry. QI-PWO and QI-PWO+ are almost unchanged by the coil-feasibility surrogate and remain comparable to the \textit{qh} cases.
This result does not establish \textit{qi-pwO} as a general replacement for \textit{qi}. Rather, preserving the weak-field-side mapping while relaxing the strong-field-side requirement provides additional design freedom that may be useful for improving coil feasibility. 
Broader configuration studies are needed to determine how generally this potential can be realized.

\subsection{Plasma boundary geometries}
\label{sec:boundary_geometry}

\begin{figure*}[htbp]
  \centering
  \begin{minipage}[t]{0.49\textwidth}
    \centering
    \includegraphics[width=\linewidth]{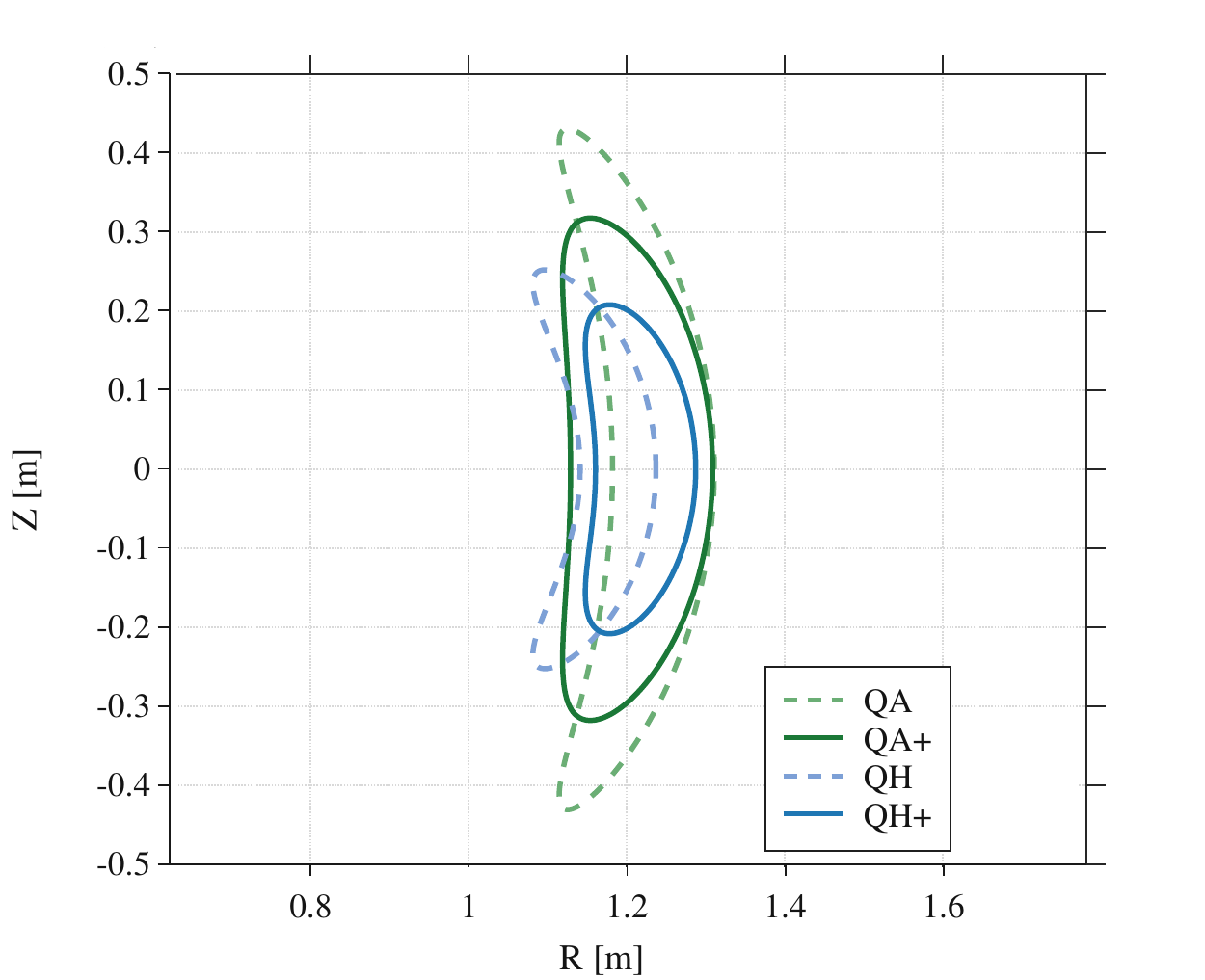}
    \smallskip
    (a) qa and qh
  \end{minipage}\hfill
  \begin{minipage}[t]{0.49\textwidth}
    \centering
    \includegraphics[width=\linewidth]{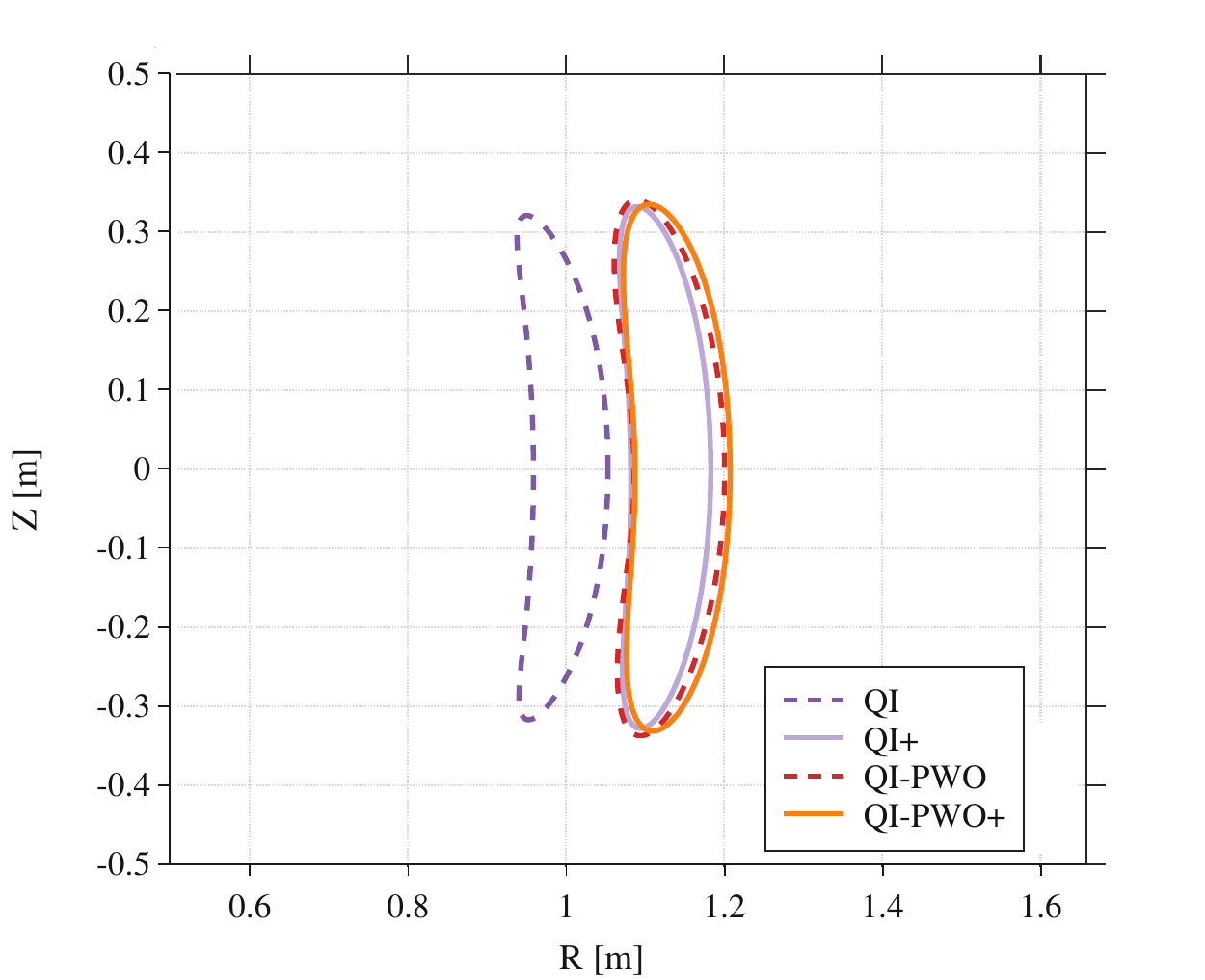}
    \smallskip
    (b) qi and qi-pwO
  \end{minipage}
  \caption{Bean-shaped cross-sections of the LCFS for the optimized configurations.}
  \label{fig:plasma_boundary}
\end{figure*}

Figure \ref{fig:plasma_boundary} shows that QSS-optimized configurations generally have less elongated boundary shapes, as listed in table \ref{tab:equilibrium_params}. The maximum elongation decreases from 6.9 to 3.5 for QA, from 4.5 to 2.5 for QH, from 6.6 to 6.1 for QI, and from 5.4 to 5.1 for QI-PWO. The mean elongation follows the same trend. The minimum \(L_{\nabla B}\) also increases for QH, QI, and QI-PWO, indicating a larger estimated coil-plasma spacing in these cases.
Two details are worth noting. First, QA+ has lower elongation but a slightly smaller minimum \(L_{\nabla B}\) than QA, showing that this pre-evaluation metric does not fully determine the final coil-plasma distance. Second, the QI+ cross-section becomes close to those of QI-PWO and QI-PWO+, consistent with the convergence of their \(\iota\) profiles.

\subsection{Surface current distribution and coil manufacturability}
\label{sec:surface_current_manufacturability}

\begin{figure*}[htbp]
  \centering
  \includegraphics[width=0.8\textwidth]{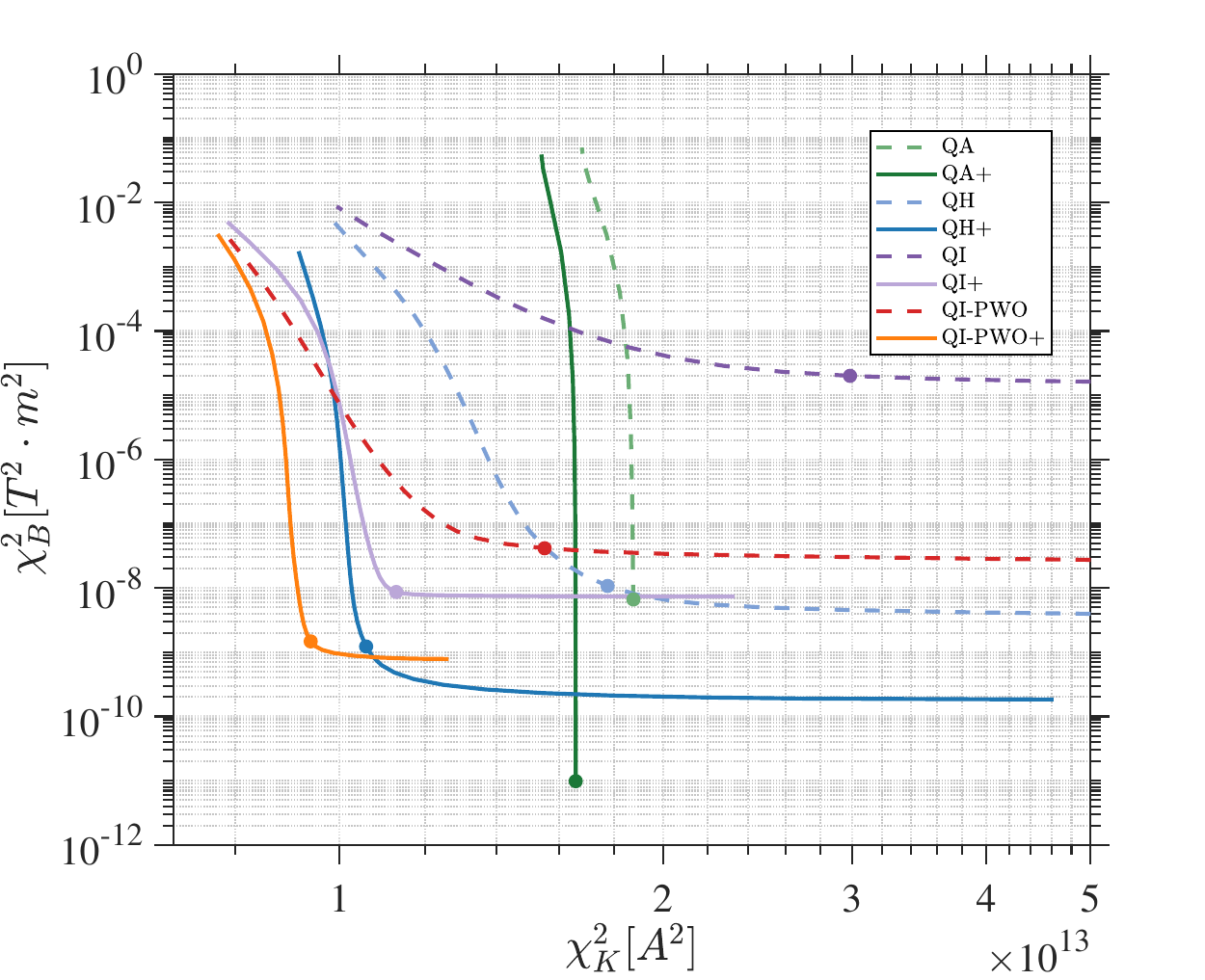}
  \caption{Regularization curves of the surface current for all 8 configurations, calculated by REGCOIL. The horizontal axis is the surface-current-density squared integral $\chi^2_K$ on the winding surface, and the vertical axis is the normal magnetic field error squared integral $\chi^2_B$ on the plasma surface $S$. The visually selected diagnostic point in the turning region of each curve is marked with a solid dot, and the endpoint corresponds to the regularization parameter $\lambda=0$.}
  \label{fig:regularization_curves}
\end{figure*}

\begin{figure*}[htbp]
  \centering
  \includegraphics[width=\textwidth]{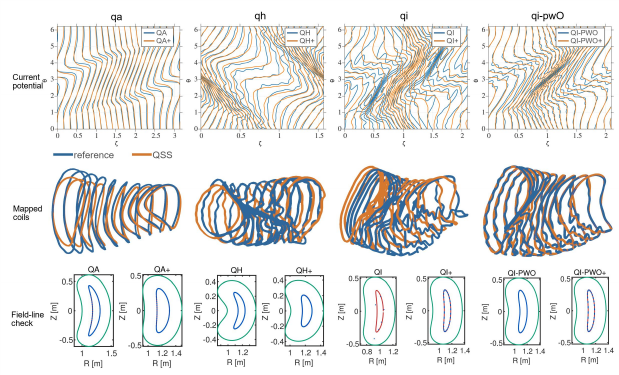}
  \caption{Full-chain verification of surface-current reconstruction, mapped coil geometry, and flux-surface recovery at the selected L-curve diagnostic point. Each column corresponds to one symmetry class. The first row shows 20 current-potential contours over one field period. The second row shows the coil shapes corresponding to 10 contours over half a field period. The third row shows Poincaré field-line checks. In the first two rows, blue and orange denote the reference and QSS-optimized configurations, respectively. In the Poincaré row, green denotes the winding-surface boundary, red denotes the target LCFS, and blue denotes traced field-line points.}
  \label{fig:coil_verification_chain}
\end{figure*}

Surface-current distributions resemble the coil shapes.
We use a representative point in the turning region of each L-curve as a diagnostic design point, as shown in Figure~\ref{fig:regularization_curves}.
The current-potential contours, mapped coil shapes, and Poincaré plots are plotted in Figure~\ref{fig:coil_verification_chain}.
In Figure~\ref{fig:regularization_curves}, the constrained cases move toward lower \(\chi_B^2\) and lower \(\chi_K^2\) at the selected turning-region diagnostic point. For QA+, \(\chi_B^2\) decreases from \(6.5\times10^{-9}\) to \(9.8\times10^{-12}\), while \(\chi_K^2\) decreases from \(1.9\times10^{13}\) to \(1.7\times10^{13}\). For QH+, the corresponding values change from \(5.6\times10^{-9}\) to \(4.8\times10^{-10}\) and from \(2.2\times10^{13}\) to \(1.1\times10^{13}\). For QI+, they change from \(2.0\times10^{-5}\) to \(7.9\times10^{-9}\) and from \(3.0\times10^{13}\) to \(1.2\times10^{13}\). QI-PWO+ also reduces both quantities relative to QI-PWO. Thus, the \(b_n^{\mathrm{max}}\) constraint improves field reconstruction and lowers the selected surface-current complexity proxy in all four paired comparisons.

Figure \ref{fig:coil_verification_chain} shows the same comparison visually. The QSS-optimized cases have smoother current-potential contours and fewer local folds. The reference QI case is the most distorted and contains a locally closed contour, while QI+ removes this feature. The mapped coil curves follow the same trend: the constrained cases are smoother and more regular, especially for QI+.
The Poincaré plots provide the field-reconstruction check. Except for reference QI, the traced flux surfaces agree well with the target LCFS. The failure of reference QI and the improvement in QI+ show why coil feasibility must be considered before the final coil-design stage.

\subsection{3D modular coils designed by FOCUS}
\label{sec:focus_coils}

\begin{figure*}[htbp]
  \centering
  \includegraphics[width=0.95\textwidth]{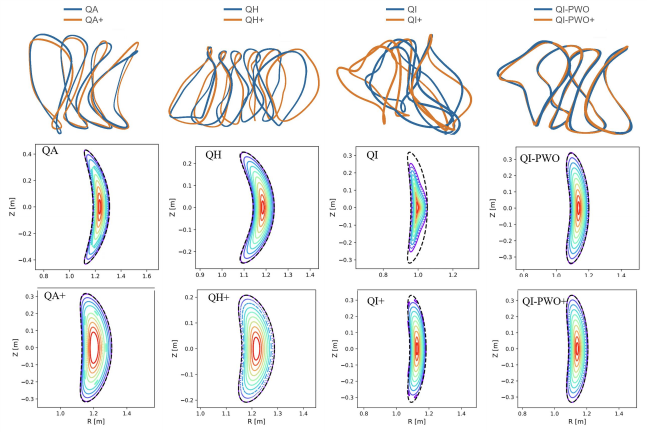}
  \caption{Three-dimensional modular coils designed by FOCUS and corresponding Poincaré cross-sections for all eight configurations. The top row shows half-period 3D coil geometries for the four symmetry classes, with blue denoting reference configurations and orange denoting QSS-optimized configurations. The middle and bottom rows show Poincaré cross-sections for the reference and optimized configurations, respectively. In these cross-sections, the colored curves are traced flux surfaces and the black dashed curve denotes the target LCFS.}
  \label{fig:focus_coils}
\end{figure*}

Now, we can design modular coils to validate whether the coil complexity is reduced.
A good coil set needs to balance physics accuracy and engineering complexity.
When comparing the QSS-optimized equilibrium with the reference case, we ensure that engineering metrics are \textit{not} worse while reducing the normal field error as much as possible.
Table \ref{tab:focus_coil_params} and Figure \ref{fig:focus_coils} summarize the FOCUS coil-design checks. These results are interpreted only as paired reference-versus-QSS comparisons within each symmetry class, because coil number, current scale, and final soft-target settings differ across classes.

The maximum curvature remains close to its FOCUS soft target in each paired comparison. The average curvature decreases from \(4.17\) to \(3.93\,\mathrm{m}^{-1}\) for QH+, from \(2.85\) to \(2.60\,\mathrm{m}^{-1}\) for QI+, and from \(2.99\) to \(2.89\,\mathrm{m}^{-1}\) for QI-PWO+, corresponding to reductions of about 5.8\%, 8.8\%, and 3.3\%. QA+ is nearly unchanged because QA's coils already have low average curvature. The coil shapes in Figure \ref{fig:focus_coils} show the same trend visually, with the clearest improvement in QI+. Coil--coil and coil--plasma distances also increase slightly for QH+ and QI-PWO+.

With respect to magnetic field reconstruction accuracy, the FOCUS area-averaged absolute relative normal-field error \(\left\langle |b_n| \right\rangle_S\) on the LCFS is lower for all QSS-optimized configurations than for their corresponding reference counterparts. It decreases from \(5.33\times10^{-4}\) to \(4.3\times10^{-4}\) for the \textit{qa} class, from \(1.51\times10^{-3}\) to \(5.51\times10^{-4}\) for the \textit{qh} class, from \(3.3\times10^{-2}\) to \(1.4\times10^{-2}\) for the \textit{qi} class, and from \(1.99\times10^{-3}\) to \(1.8\times10^{-3}\) for the \textit{qi-pwO} class. The corresponding reductions are approximately 19\%, 64\%, 58\%, and 10\%, respectively. 
Poincaré plots from filamentary coils in the second and third rows of Figure \ref{fig:focus_coils} further verify the reconstruction quality from the perspective of flux-surface integrity. For the reference QI configuration, the designed coils fail to reproduce complete closed flux surfaces, accompanied by severe edge islands and field-line diffusion. In contrast, the QSS-optimized QI+ configuration yields substantially improved flux-surface closure. For other classes, closed nested flux surfaces are well maintained in both reference and optimized cases, while the optimized configurations show better agreement between the LCFS and the target equilibrium boundary, with suppressed edge island structures. These observations are consistent with the reduction in \(\left\langle |b_n| \right\rangle_S\) shown in Table \ref{tab:focus_coil_params}, indicating that introducing the \(b_n^{\mathrm{max}}\) constraint at the stage of equilibrium optimization can improve the magnetic field reconstruction accuracy of subsequent coil design with coils of equivalent or even reduced complexity.
The FOCUS results support the effectiveness of the \(b_n^{\mathrm{max}}\) surrogate: adding this term into equilibrium optimization will help find configurations that have favourable physics properties and feasible coils. 

\begin{table}[htbp]
  \centering
  \caption{Achieved parameters of 3D modular coils designed by FOCUS for all eight configurations. The geometric soft targets are specified in Section~\ref{sec:focus}; deviations from those targets are allowed. The last row reports the area-averaged absolute relative normal-field error \(\left\langle |b_n| \right\rangle_S\) on the LCFS.}
  \label{tab:focus_coil_params}
  \resizebox{\linewidth}{!}{
  \begin{tabular}{lcccccccc}
    \toprule
    Parameter & QA & QA+ & QH & QH+ & QI & QI+ & QI-PWO & QI-PWO+ \\
    \midrule
    average length (m) & 5.00 & 5.00 & 2.10 & 2.10 & 4.0 & 4.0 & 3.50 & 3.50 \\
    maximum curvature ($\text{m}^{-1}$) & 9.74 & 9.74 & 9.6 & 9.6 & 9.4 & 9.4 & 9.02 & 9.01 \\
    average curvature ($\text{m}^{-1}$) & 1.69 & 1.69 & 4.17 & 3.93 & 2.85 & 2.6 & 2.99 & 2.89 \\
    coil-coil distance (m) & 0.115 & 0.116 & 0.059 & 0.061 & 0.034 & 0.034 & 0.05 & 0.05 \\
    coil-plasma distance (m) & 0.31 & 0.312 & 0.12 & 0.12 & 0.13 & 0.13 & 0.157 & 0.158 \\
    $\left\langle |b_n| \right\rangle_S$ & $5.33\times10^{-4}$ & $\mathbf{4.3\times10^{-4}}$ & $1.51\times10^{-3}$ & $\mathbf{5.51\times10^{-4}}$ & $3.3\times10^{-2}$ & $\mathbf{1.4\times10^{-2}}$ & $1.99\times10^{-3}$ & $\mathbf{1.8\times10^{-3}}$ \\
    \bottomrule
  \end{tabular}
  }
\end{table}

\section{Discussion and conclusions}
\label{sec:conclusion}

This work develops a QSS optimization workflow that embeds a coil-feasibility surrogate into plasma-boundary optimization. The \(b_n^{\mathrm{max}}\) computed by the surface current on a uniformly offset winding surface penalizes the coil feasibility. Across various examples covering different magnetic symmetries, the constraint reduces field-reconstruction error, smooths plasma and winding-surface geometry, and lowers the selected surface-current complexity proxy, while retaining favourable magnetic symmetry and transport properties. The method provides a proof-of-principle route for balancing confinement quality and coil feasibility.

The main findings are: (1) \(b_n^{\mathrm{max}}\) can be used as a compact coil-feasibility surrogate inside equilibrium optimization; (2) the same workflow can treat qa, qh, qi, and qi-pwO targets; and (3) QI-PWO+ gives the best compromise among the qi-type cases studied here, illustrating the potential superiority of the qi-pwO route for engineering-oriented optimization.

Several limitations remain. First, a rigorous statistical relation between \(b_n^{\mathrm{max}}\) and filamentary-coil complexity is beyond the scope of this work; it would require a large database of equilibria, winding surfaces, REGCOIL current-potential solutions, and optimized coils. The present study provides case-wise evidence only. Second, all calculations are fixed-boundary, zero-beta, and vacuum. Future work should extend the workflow to finite beta, bootstrap current, MHD stability, and turbulence transport. A database study will also be needed to test whether \(b_n^{\mathrm{max}}\), elongation, \(L_{\nabla B}\), surface-current metrics, and filamentary-coil metrics show robust statistical trends across broader configuration sets.

\section*{Data availability statement}
The equilibrium files, coil files, and source data supporting the figures will be deposited in Zenodo and made publicly available upon publication. During peer review, these data are available from the corresponding author upon reasonable request. The source code developed specifically for this study is not publicly available; the calculations use the cited community codes SIMSOPT, VMEC, REGCOIL, OOPS, and FOCUS.

\section*{Acknowledgments}
This work was supported by the National Natural Science Foundation of China (NSFC) under Grant Nos.~12405267 and 12475229, and the Strategic Priority Research Program of the Chinese Academy of Sciences under Grant No.~XDB0790302.

\newcommand{\newblock}{}
   \bibliographystyle{unsrt}
\bibliography{ref}

\end{document}